\documentstyle[prl,aps,epsf]{revtex}
\begin{document}
\draft

\twocolumn[\hsize\textwidth\columnwidth\hsize\csname@twocolumnfalse\endcsname

\title{Hall noise and transverse freezing 
in driven vortex lattices}

\author{Alejandro B. Kolton and Daniel Dom\'{\i}nguez}
\address{Centro At\'{o}mico Bariloche, 8400 S. C. de Bariloche,
Rio Negro, Argentina}
\author{Niels Gr{\o}nbech-Jensen}
\address{Theoretical Division, Los Alamos National Laboratory,
Los Alamos, NM 87545, USA}

\date{\today}
\maketitle
\begin{abstract}
We study driven vortices lattices in
superconducting thin films. 
Above the critical force $F_c$ we find two  dynamical phase
transitions at $F_p$ and $F_t$, which could be observed in
simultaneous noise measurements of the longitudinal and the Hall voltage.
At $F_p$ there is a transition
from plastic flow to smectic flow
where  the voltage noise is isotropic (Hall noise = longitudinal
noise) and there is a peak in the differential resistance.
At $F_t$ there is a sharp transition to a frozen transverse solid
where the Hall noise falls down abruptly
and vortex motion is localized in the transverse direction. 

\end{abstract}


]                

\narrowtext

The study of the collective motion of vortex lattices in superconductors
has brought new concepts in the non-equilibrium statistical
physics of driven disordered media
\cite{KV,plastic,filam,olson,GLD,bmr,scheidl,%
bhatta,hellerq,yaron,pardo,%
moon,ryu,spencer,olson2,dd,dgb,aranson}.
The prediction \cite{KV} of a {\it dynamical phase transition}
upon increasing drive, from
a fluidlike plastic flow  regime \cite{plastic,filam,olson} 
to a coherently moving solid \cite{KV}, 
 has motivated an outburst
of recent theoretical \cite{GLD,bmr,scheidl}, 
experimental \cite{bhatta,hellerq,yaron,pardo}, and simulation
\cite{moon,ryu,spencer,olson2,dd,dgb,aranson} work. 
The relevant physics of the
high velocity driven phase is controlled by the transverse displacements
(in the direction perpendicular to the
driving force) \cite{GLD}, leading to a new class of
driven systems characterized by {\it anisotropic spatial structures} 
with transverse periodicity \cite{GLD,bmr,scheidl}.
Recently, these moving anisotropic vortex structures have been observed 
experimentally by Pardo {\it et al.} \cite{pardo}, and their different
features have been studied  in 2D \cite{moon,ryu,spencer,olson2,dd} and 
3D \cite{dgb,aranson} simulations.
In this letter, we show that a better insight on the
moving phases can be obtained from studying the
{\it anisotropic temporal fluctuations}. 
We find
two dynamical phase transitions which could be observed 
experimentally by
measuring voltage noise \cite{yeh,noise} 
both in the longitudinal and transversal directions. 
In the transverse
direction, where the system is {\it not driven},
we can study diffusion, from which we 
 define an effective temperature in analogy with equilibrium physics
\cite{KV,scheidl}.

The equation of
motion of a vortex in position ${\bf r}_i$ is:
\begin{equation}
\eta \frac{d{\bf r}_i}{dt} = -\sum_{j\not= i}{\bf\nabla}_i U_v(r_{ij})
-\sum_p{\bf \nabla}_i U_p(r_{ip}) + \bf{F},
\end{equation} 
where $r_{ij}=|{\bf r}_i-{\bf r}_j|$ is the distance between vortices $i,j$,
$r_{ip}=|{\bf r}_i-{\bf r}_p|$ is the distance between the vortex $i$ and
a pinning site at ${\bf r}_p$, $\eta=\frac{\Phi_0H_{c2}d}{c^2\rho_n}$ is the
Bardeen-Stephen friction and ${\bf F}=\frac{d\Phi_0}{c}{\bf J}\times{\bf z}$
is the driving force due to an applied current ${\bf J}$.
A two-dimensional superconductor is realized in 
thin films of thickness $d$ where $d\ll\lambda$, which have an effective
penetration depth $\Lambda=2\lambda^2/d$. Since $\Lambda$ is of the order
of the sample size ($\Lambda\approx200\mu m$ in \cite{hellerq}),
 the vortex-vortex interaction is logarithmic:
$U_v(r)=-A_v\ln(r/\Lambda)$, with $A_v=\Phi_0^2/8\pi\Lambda$
\cite{filam,ryu}.
The vortices interact with a random uniform distribution of
attractive pinning centers with 
$U_p(r)=-A_p e^{-(r/\xi)^2}$ with $\xi$ being the coherence length. 
We normalize length scales by $\xi$, energy scales by $A_v$, 
and time is normalized by 
$\tau=\eta\xi^2/A_v$.  We consider $N_v$ vortices and $N_p$ pinning
centers in a rectangular box of size $L_x\times L_y$, 
and the normalized  vortex density is $n_v=N_v\xi^2/L_xL_y=B\xi^2/\Phi_0$.
Moving vortices induce a total electric field  ${\bf
E}=\frac{B}{c}{\bf v}\times{\bf z}$, with ${\bf v}=\frac{1}{N_v}\sum_i 
{\bf v}_i$.

We study the dynamical regimes in the velocity-force (voltage-current)
characteristics at $T=0$, solving   Eq.(1) for increasing values of 
 ${\bf F}=F{\bf y}$. 
We consider a constant vortex density $n_v=0.12$ in
a box with $L_x/L_y=\sqrt{3}/2$, and 
$N_v=64,144,196,256,400,784$ (we show results for $N_v=400$). 
We take a pinning strengh of $A_p/A_v=0.2$
with a density of pinning centers $n_p=0.24$. We use periodic boundary
conditions and the periodic long-range logarithmic interaction is evaluated with 
an exact and fast converging sum \cite{log}. 
The equations are integrated
with a time step of $\Delta t=0.01\tau$ and averages are
evaluated in $32768$ integration steps after $2000$ iterations for 
equilibration (when the total energy reaches a stationary value). 
Each simulation is started at $F=0$ with an ordered
triangular lattice and slowly increasing the
force in steps of $\Delta F= 0.05$ up to values as high as $F=8$.

First we start  by looking at the vortex trajectories and their 
translational order in the steady state
phases as was done in \cite{moon,ryu,spencer,olson2,dd,dgb,aranson}.
In Figure 1(a-c) we show the vortex trajectories $\{ {\bf r}_i(t)\}$ 
for typical
values of $F$ by plotting all the positions of the vortices for all the
time iteration steps. 
We also study the 
time-averaged structure factor 
$S({\bf k})= \langle|\frac{1}{N_v}\sum_i \exp[i{\bf k}\cdot{\bf
r}_i(t)]|^2\rangle$, which is shown in  Fig.  1(d-f).
In Fig. 2(a) we plot  the average vortex velocity 
$V=\langle V_y(t)\rangle=\langle\frac{1}{N_v}\sum_i \frac{dy_i}{dt}\rangle$,
 in the 
direction of the force as a function of $F$ and its corresponding
derivative $dV/dF$.  
($V=E/\rho_fJ_0$ with $\rho_f$
the flux flow resistivity  and $J_0=cA_v/d\xi\Phi_0$,
therefore $dV/dF$ is proportional to
the differential resistivity
$dV/dF=\rho_f^{-1} dE/dJ$).
Below a critical force $F_c \approx 0.25$ all the vortices are
pinned and 
\begin{figure}
\centerline{\epsfxsize=8.5cm \epsfbox{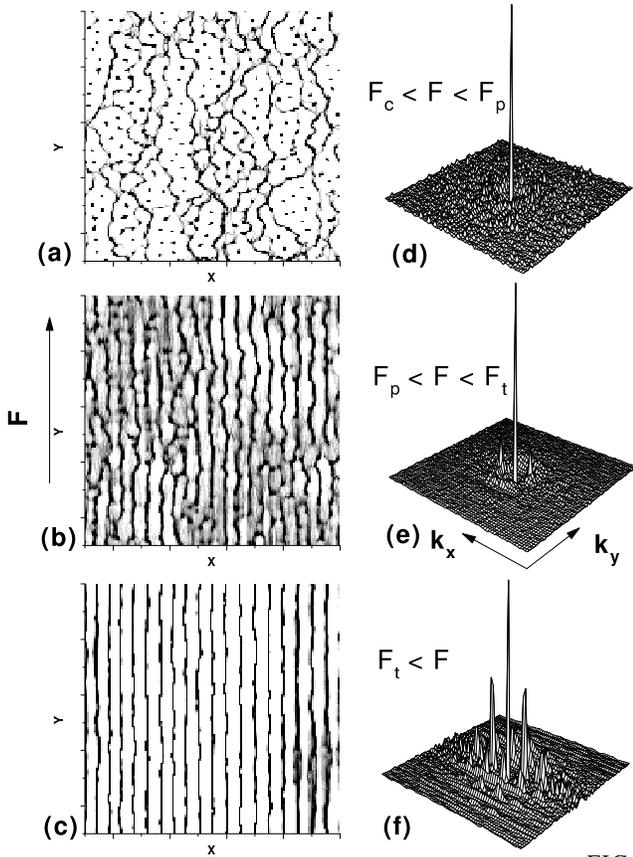}}
\caption{
Vortex trajectories: (a) $F=0.5$, (b) $F=2.5$, (c) $F=7$.
Surface intensity plot of the averged structure factor $S({\bf k})$:
(d)$F=0.5$, (e) $F=2.5$, (f) $F=7$.}
\end{figure}\noindent
there is no motion, $V=0$. Above $F_c$, we distinguish
between three different dynamical regimes:

{\it (i) Plastic flow}: $F_c<F<F_p$.
At $F_c$ vortices start to move in 
a few filamentary channels, as was also seen in \cite{filam}.
A typical situation is shown in Fig. 1(a), where 
a fraction of the
vortices are moving in an intrincate network of channels. 
As the force is increased a higher fraction of  vortices
is moving.
In this regime, vortices can move in the transverse direction
(perpendicular to ${\bf F}$) through the tortuous structure
of channels \cite{olson}.
We see 
in Fig. 1(d) that  $S({\bf k})$ only has the central peak showing
the absence of ordering in this plastic flow regime \cite{plastic,filam,olson}.

{\it (ii) Smectic flow}: $F_p<F<F_t$.
We observe a peak in the
differential resistance at a characteristic force $F_p\approx 1.3$.
At $F=F_p$ we see that
{\it all} the vortices are moving in a 
seemingly isotropic channel network 
with maximum interconnectivity.
In other simulations
the peak in the differential resistance was found to coincide
with a maximum in the number of defects \cite{olson2} and
with the onset of orientational order \cite{ryu}.
Also, the value of $F_p$ was taken
in the experiment of Hellerqvist {\it et al} \cite{hellerq} 
as an indication of
a dynamical phase transition.
In fact, we 
find that above $F_p$ a new dynamic regime sets in. In this 
case, as we show in Fig. 1(b), all the vortices are moving
\begin{figure}
\centerline{\epsfxsize=8.5cm \epsfbox{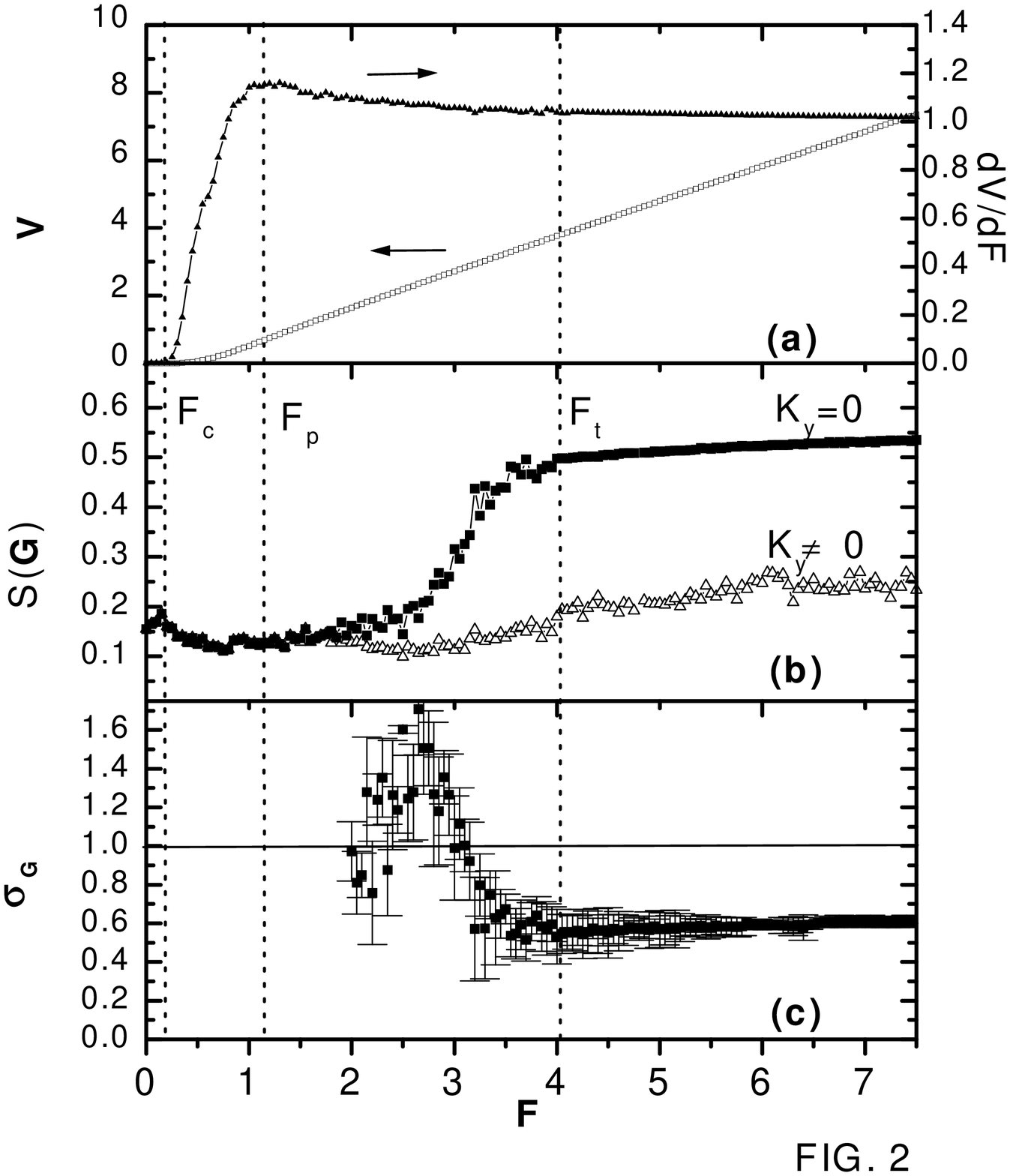}}
\caption{(a) Velocity-force curve (voltage-current characteristics),
left scale, white points. $dV/dF$ curve (differential resistance),
right scale, black points.
(b) Intensity of the Bragg peaks.
For smectic ordering, $S(G_1)$, $K_y=0$ : black points.
For longitudinal ordering, $S(G_{2,3})$, $K_y\not=0$ : white points.
(c) Finite size exponent $\sigma_G$
($S(G_1)\sim N_v^{-\sigma_G}$).}
\end{figure}\noindent 
in trajectories that are mostly parallel to the force, 
forming ``elastic channels''. 
Two small Bragg peaks appear in the
structure factor along the $K_y=0$ axis, as seen in Fig. 1(e),
which correspond to ${\bf G}_1=(\pm 2\pi/a_0,0)$. 
This is consistent with the onset of ``smectic'' ordering
\cite{bmr,moon}
 in the transverse direction with elastic channels separated by 
 a distance  $\sim a_0 = n_v^{-1/2}$.
In this regime the transverse motion consists in
vortex jumps from one channel to another, resembling
``thermally'' activated transitions induced by local chaos. 
The rate of these transitions
decreases with increasing force, and they correspond
to the permeation modes discussed in \cite{bmr}. 
In Fig. 2(b) we plot the magnitude 
of the Bragg peaks at $G_1$, $S_s=S(G_1)$, corresponding to
smectic ordering ($K_y=0$), and the other neighbouring peaks
at ${\bf G}_2=\pm2\pi/a_0(1/2,\sqrt{3}/2)$ and
${\bf G}_3=\pm2\pi/a_0(-1/2,\sqrt{3}/2)$
, $S_{l}=(S(G_2)+S(G_3))/2)$, corresponding
to longitudinal ordering ($K_y\not=0$).
We see that above $F_p$ the 
intensity of the smectic peak $S_s$  starts to grow and $S_s\gg S_l$,
while below $F_p$ the spatial structure is isotropic, $S_s=S_l\ll 1$.
The Bragg peak heights depend with system size 
as $S(G) \sim N_v^{-\sigma_G}$, where
$\sigma_G=0$ means long-range order (LRO), $0<\sigma_G<1$ means
quasi long-range order (QLRO) and $\sigma_G=1$ means short-range order
(SRO). In Fig. 2(c) we show the corresponding results 
for the smectic peak $S_s=S(G_1)$ for sizes $N_v=256,400,784$.
We see that $\sigma_G \gtrsim 1$ in this regime: there is only
short range smectic order,
thus this phase corresponds to a liquid. 
In this sense the
transition at $F_p$ is a dynamic transition in the flow, as found in
\cite{dd} for strong disorder ($n_p=1$).
When approaching $F_t$ 
(see below) 
we see a precursor of QLRO 
($\sigma_g<1$) but with strong fluctuations.

{\it (iii) Frozen transverse solid}: $F>F_t$.
At a new characteristic force $F_t$, the 
jumps between channels suddenly stop and vortex motion becomes
frozen in the direction perpendicular to ${\bf F}$. An example for
$F>F_t$ is shown in Fig. 1(c) where we see well defined elastic channels 
parallel to ${\bf F}$. 
The corresponding structure factor is
in Fig. 1(e) where
new peaks appear in $S({\bf k})$ in directions
with $K_y\not=0$, like $G_2$, $G_3$, showing  that there is some
 longitudinal ordering between the channels.  These
 extra peaks are smaller than the smectic peaks, and $S({\bf k})$ is 
 very anisotropic. 
We note in Fig. 2(a) that $F_t$ is the point where the
noisy behavior in $dV/dF$ ceases.
A simialar criterion was used by Bhattacharya and Higgins to define
their dynamical phase diagram \cite{bhatta}.
In Fig. 2(b) we see that in $F_t$
there is an increase in the longitudinal ordering
$S_l$, and both $S_s$ and $S_l$ tend to saturate at an almost constant value
for $F\gg F_t$. In  Fig. 2(c) 
we find that there is smectic QLRO \cite{GLD,bmr,moon}
with a value of $0.5<\sigma_G <0.7$. 
However, we were
not able to obtain a reliable finite size analysis for the
longitudinal peaks since
they have large fluctuations from size to size.
Also, 
we find hysteresis in $S_s$
around $F_t$ when decreasing $F$.

 A better understanding of the {\it dynamical} transitions
can be obtained from studying the temporal behavior of the system in
both directions. It has been observed experimentally
 that the longitudinal voltage can show low frequency noise 
 \cite{yeh,noise}.
  This voltage noise reaches a very large
 value above the critical current, which
 has been attributed to plastic flow \cite{noise}, and then
 the  noise decreases for large current.
In addition, even when the total dc transverse
voltage $\langle V_x\rangle=
\langle\frac{1}{N_v}\sum_i \frac{dx_i}{dt}\rangle$ is zero, 
it can also have fluctuations and noise \cite{niels}. 
In fact, it is easy to understand that this {\it Hall noise}
 will be closely related to the wandering and wiggling of
the plastic flow channels and to the jumps between elastic channels
in the smectic phase.
We have calculated the
power spectrum of both the longitudinal voltage, 
$S_y(f)=|\frac{1}{T}\int_0^Tdt (V_y(t)-V)\exp(i2\pi ft)|^2$,
and the transverse voltage, 
$S_x(f)=|\frac{1}{T}\int_0^Tdt  V_x(t)\exp(i2\pi ft)|^2$.
The low frequency noise is defined as
 $P_{x,y}=\lim_{f\rightarrow0}S_{x,y}(f)$.
In Fig. 3(a) we show the values of the longitudinal noise $P_y$ and
the Hall noise $P_x$ as a function of the force 
($P_{x,y}$ were approximated from the average of the 10 lowest frequencies).
We see that near the critical force, the 
longitudinal noise is large while
the Hall noise is one order 
\begin{figure}
\centerline{\epsfxsize=8.5cm \epsfbox{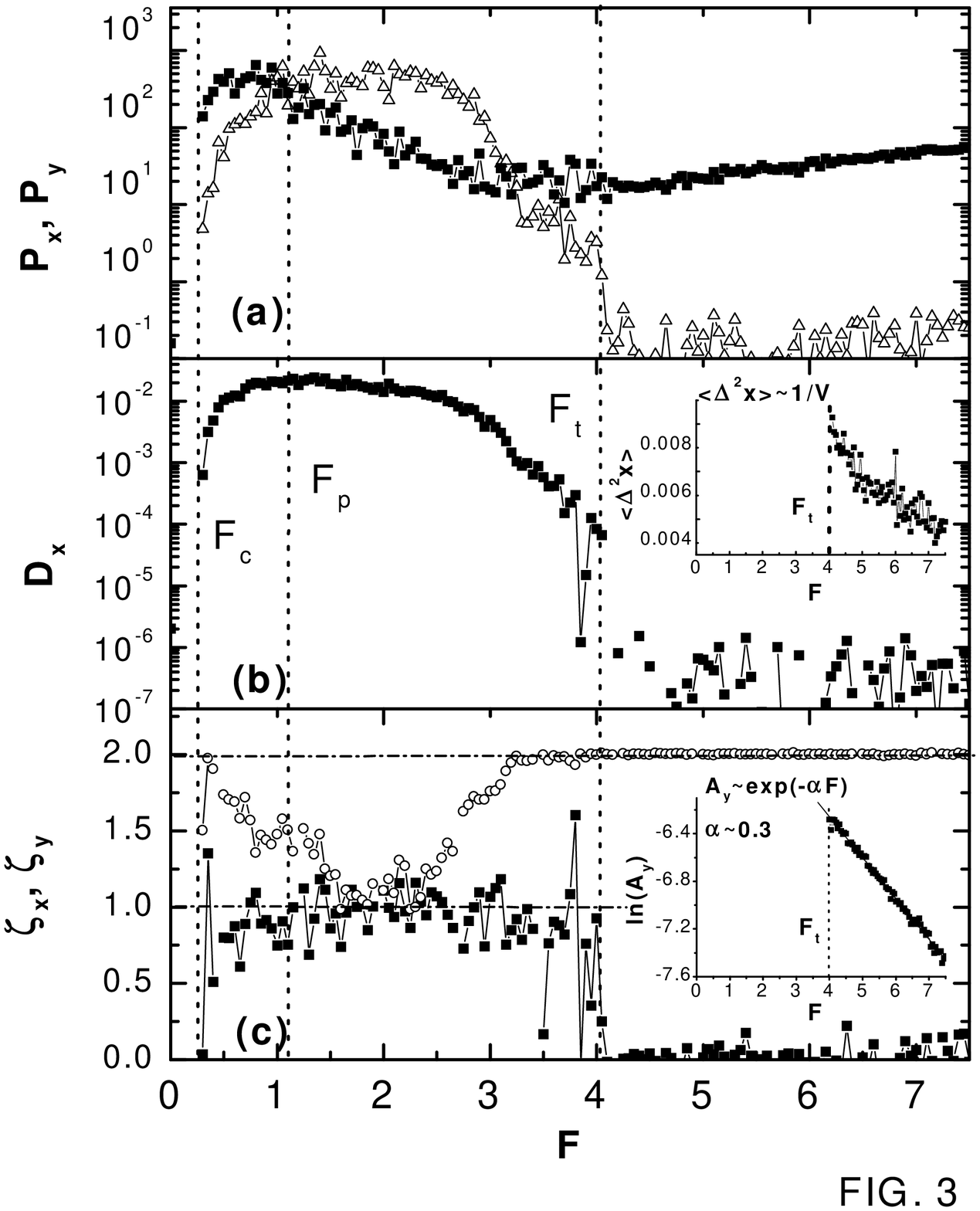}}
\caption{(a) Low frequency voltage noise vs $F$.
Longitudinal voltage noise, $P_y$, black points. 
Hall noise, $P_x$, white points.
(b) Diffusion coefficient for transversal motion, $D_x$.
The inset shows the average transverse quadratic displacement
$\langle\Delta^2 x\rangle$ in the frozen phase, $F>F_t$.
(c) Diffusion exponents $\zeta_x$ (black dots) and $\zeta_y$ (white dots)
defined from:
$\langle[\tilde{x}(t)-\tilde{x}(0)]^2\rangle\sim t^{\zeta_x}$,
$\langle[\tilde{y}(t)-\tilde{y}(0)]^2\rangle\sim t^{\zeta_y}$.
The inset shows a plot of $A_y$ defined from a fit 
$\langle[\tilde{y}(t)-\tilde{y}(0)]^2\rangle=A_y t^2$ for $F>F_t$. }
\end{figure}\noindent
of magnitude smaller. In this regime of plastic flow the
noise is dominated by fluctuations in the direction of motion
and by channel tortuosity \cite{olson},
and since there are few channels the Hall noise is small. When the number
of channels increases 
the vortices wander more in the $x$
direction and the Hall noise increases. 
At $F_p$ the voltage noise becomes isotropic, $P_x=P_y$. 
This is the point where we have seen
the highest interconnection in the channel network.
Above $F_p$, the onset of elastic channels and smectic ordering reduces the
longitudinal noise. On the other hand, the Hall noise remains large 
due to the ``activated'' 
jumps between elastic channels. 
At $F_t$ the Hall noise falls abruptly, nearly two orders of magnitude.
This corresponds to a {\it freezing transition} 
of vortex motion in the transverse
direction. Above $F_t$ there are no more vortex jumps between elastic
channels. The low frequency noise can be closely related to diffusive
motion for large times. We analyze the average 
quadratic displacements of vortices in both directions
from their center of mass 
position $(X_{cm}(t),Y_{cm}(t))$ as a function of time. We  define 
$w_x(t)=\frac{1}{N_v}\sum_i[\tilde{x}_i(t)-\tilde{x}_i(0)]^2$, and
$w_y(t)=\frac{1}{N_v}\sum_i[\tilde{y}_i(t)-\tilde{y}_i(0)]^2$
where $\tilde{x}_i(t)=x_i(t)-X_{cm}(t)$, $\tilde{y}_i(t)=y_i(t)-Y_{cm}(t)$.
We observe that the vortex motion for $F_c < F < F_t$ is diffusive
in the transverse direction, $w_x(t)\sim D_x t$.
In Fig. 3(b) we show the behavior of the
transverse diffusion coefficient $D_x$. It closely follows the
behavior of the Hall noise, $D_x \propto P_x$. 
The transverse
diffusion is maximum at the peak in the differential resistance, $F_p$,
and it has an abrupt jump to zero at $F_t$ indicating the transverse 
freezing transition. It is interesting to note that melting transtions
also show a jump in the diffusion coefficient.
Above $F_t$, the transverse wandering is independent of
time since vortex motion is
localized in the $x$ direction:
 $w_x(t) \approx \langle\Delta^2 x\rangle$.
In the inset of Fig. 3(b) we show 
$\langle\Delta^2x\rangle$  vs. $F$ for $F>F_t$. We find that
$\langle\Delta^2x\rangle\approx 0.02 a^2$ at $F_t$, 
consistent with a Lindemann criterion for melting \cite{scheidl}.
Since in equibrium the diffusion coefficient $D$ is proportional to $T$,
we interpret $D_x$ as an indication of an ``effective temperature''
$T_{eff}$.
This means that $T_{eff}$ raises above the critical force,
has a maximum at the peak in the differential resistance and  decreases
when the motion starts to order. 
Following the same idea, above $F_t$ we can interpret
$\langle\Delta^2x\rangle$
as proportional to $T_{eff}$. Here we find that $\langle\Delta^2x\rangle$
 decreases with $F$, consistent with 
 a $T_{eff}\sim 1/V$ behavior \cite{KV,scheidl}. 

The long-time behavior is better understood by studying the diffusion
exponents $w_{x}(t)\sim  t^{\zeta_{x}}$ and 
$w_{y}(t)\sim t^{\zeta_{y}}$. In Fig. 3(c) we show the behavior
of $\zeta_{x,y}$. We see that for $F_c < F < F_t$, $\zeta_x \approx 1$
corresponding to normal diffusion, while for $F>F_t$, $\zeta_x \approx 0$
corresponding to the freezing of transverse motion. 
More interestingly, the motion in the longitudinal direction is
always superdiffusive, $\zeta_y >1$. Above $F_t$, in the frozen phase,
the longitudinal fluctuations become {\it ballistic}, $\zeta_y\approx 2$.
Since the vortex positions are localized in the $x$ direction,
we see that $w_y(t)=\langle (y(t)-y(0))^2\rangle  \approx 
A_y t^2$, with $A_y=\langle \Delta^2 v_y\rangle$ the dispersion of the
average velocities of the elastic channels. In the inset
of Fig. 3(c) we find that $A_y$
decreases exponentially with the force $A_y\sim \exp(-\alpha F)$.
[This suggests a ``thermal'' distribution of channel velocities,
since $T_{eff}\sim 1/F$, then $\langle\Delta^2 v_y\rangle \sim \exp(-\alpha'/
T_{eff})$]. In the smectic flow 
region  an exponent $1<\zeta_y<2$
can be explained from assuming  that vortices move most of the
time in the elastic channels and occasionally have diffusive 
jumps between channels in the transverse direction.

In conclusion, we have obtained evidence
of two dynamical phase transitions 
which can be verified experimentally
by measurements of voltage noise {\it and Hall noise}.
The first transition at $F_p$ is the point of isotropic
noise and maximum transverse diffusion (i.e., maximum effective temperature)
and corresponds to the observed peak in the differential resistance.
The second transition at $F_t$ is a freezing transition in the {\it transverse}
direction, where 
the transverse diffusion vanishes
abruptly and
the Hall noise drops many orders of magnitude.

We acknowledege discussions with
A. R. Bishop, 
P. Cornaglia, Y. Fasano, F. de la Cruz, S. Bhattacharya,  V. B. Geshkenbein,
F. Pardo, V. M. Vinokur, M. B. Weissman, and G. Zim\'{a}nyi.
This work has been supported by a grant from ANPCYT (Argentina).
D. D. and A.B.K. also acknowledge support from Fundaci\'{o}n
Antorchas, Conicet  and CNEA (Argentina).
Parts of this work were performed under
the auspices of the U.S. D.o.E.

\end{document}